\documentclass[3p,times]{elsarticle}
\usepackage{ecrc}
\volume{00}

\firstpage{1}

\journalname{}

\runauth{Y. Wang, et al.}

\jid{}

\jnltitlelogo{}

\CopyrightLine{2019}{Published by Elsevier Ltd.}

\usepackage{url}
\usepackage{multirow}
\usepackage{booktabs}
\usepackage{longtable}
\usepackage{amssymb}
\usepackage{mathrsfs}
\usepackage{graphicx}
\usepackage{amsfonts}
\usepackage{amsmath}
\usepackage{hyperref}
\usepackage{algorithm}
\usepackage{algorithmic}
\usepackage{diagbox}
\usepackage{subfig}
\usepackage{threeparttable}
\usepackage{subfig}
\usepackage{framed}
\usepackage{rotating}

\begin{document}

\begin{frontmatter}

\title{Unsupervised Machine Learning for the Discovery of Latent Disease Clusters and Patient Subgroups Using Electronic Health Records}


\author[affl1]{Yanshan Wang \corref{cor1}}
\ead{wang.yanshan@mayo.edu}

\author[affl1]{Yiqing Zhao}

\author[affl2]{Terry M. Therneau}

\author[affl2]{Elizabeth J. Atkinson}

\author[affl1]{Ahmad P. Tafti}

\author[affl1]{Nan Zhang}

\author[affl3]{Shreyasee Amin}

\author[affl4]{Andrew H. Limper}

\author[affl1]{Hongfang Liu \corref{cor1}}
\ead{liu.hongfang@mayo.edu}

\cortext[cor1]{Corresponding author}

\address[affl1]{Division of Digital Health Sciences, Department of Health Sciences Research, 
Mayo Clinic, Rochester, Minnesota, USA\\}
\address[affl2]{Division of Biomedical Statistics and Informatics, Department of Health Sciences Research, Mayo Clinic, Rochester, Minnesota, USA\\}
\address[affl3]{Division of Rheumatology, Department of Medicine, Mayo Clinic, Rochester, Minnesota, USA\\}
\address[affl4]{Division of Pulmonary and Critical Care Medicine, Department of Internal Medicine, Mayo Clinic, Rochester, Minnesota, USA\\}

\begin{abstract}
Machine learning has become ubiquitous and a key technology on mining electronic health records (EHRs) for facilitating clinical research and practice. Unsupervised machine learning, as opposed to supervised learning, has shown promise in identifying novel patterns and relations from EHRs without using human created labels. In this paper, we investigate the application of unsupervised machine learning models in discovering latent disease clusters and patient subgroups based on EHRs. We utilized Latent Dirichlet Allocation (LDA), a generative probabilistic model, and proposed a novel model named Poisson Dirichlet Model (PDM), which extends the LDA approach using a Poisson distribution to model patients' disease diagnoses and to alleviate age and sex factors by considering both observed and expected observations. In the empirical experiments, we evaluated LDA and PDM on three patient cohorts, namely Osteoporosis, Delirium/Dementia, and Chronic Obstructive Pulmonary Disease (COPD)/Bronchiectasis Cohorts, with EHR data retrieved from the Rochester Epidemiology Project (REP) medical records linkage system, for the discovery of latent disease clusters and patient subgroups. We compared the effectiveness of LDA and PDM in identifying latent disease clusters through the visualization of disease representations learned by two approaches. We also tested the performance of LDA and PDM in differentiating patient subgroups through survival analysis, as well as statistical analysis of demographics and Elixhauser Comorbidity Index (ECI) scores in those subgroups. The experimental results show that the proposed PDM could effectively identify distinguished disease clusters based on the latent patterns hidden in the EHR data by alleviating the impact of age and sex, and that LDA could stratify patients into more differentiable subgroups than PDM in terms of p-values. However, the subgroups discovered by PDM might imply the underlying patterns of diseases of greater interest in epidemiology research due to the alleviation of age and sex. Both unsupervised machine learning approaches could be leveraged to discover patient subgroups using EHRs but with different foci.

\end{abstract}

\begin{keyword}
unsupervised machine learning \sep artificial intelligence \sep electronic health records \sep epidemiology \sep aging
\end{keyword}

\end{frontmatter}

\section{Introduction}

The rapid adoption of electronic health records (EHRs) has enabled the use of the EHR data for primary and secondary purposes, such as clinical process optimization, clinical decision support, treatment outcome improvement, clinical research, and epidemiological monitoring of the nation's health~\cite{hersh2007adding}. An emerging use on the EHR data is to develop advanced machine learning models, primarily supervised learning, to discover new interconnections between diseases, facilitate precise prediction of health status, and help effectively prevent diseases or disabilities~\cite{obermeyer2016predicting,miotto2016deep,xiao2018opportunities,wang2018clinical}. As opposed to supervised learning, unsupervised machine learning has been introduced to identify new patterns and relations in the irregularly-sampled data without using human created labels~\cite{lecun2015deep,chen2018neural}, mostly for predictive modeling, such as the prediction of patient health status~\cite{miotto2016deep}, disease progression trajectory prediction~\cite{wang2014unsupervised}, or phenotypes prediction~\cite{pivovarov2015learning,son2018deep}. In this paper, we investigate the use of unsupervised machine learning in the discovery of latent disease clusters and patient subgroups using EHRs.


The problem of discovering potential disease clusters and patient subgroups is extremely important for the study of aging. According to the United Nations Population Division, the global share of older people ($\geqslant$ 60 years old) increased from 8\% in 1950 and 9\% in 1990 to 12 \% in 2013, and will continue to grow to an estimated 21\% in 2050~\cite{united2013world}. The development of chronic illness plays an important role in this demographic shift. The older people who survive with chronic illnesses are more likely to develop additional chronic illnesses~\cite{divo2014ageing}. Traditionally, most comorbidities have been studied separately ~\cite{vanfleteren2013clusters}, however, it is common for most older people to have two or more chronic morbidities. Therefore, discovering disease clusters will help in the systematic examination of all comorbidities for associations with a specific condition and improve risk assessment and future prediction. Moreover, identifying new disease clusters that may reflect underlying mechanisms (``latent traits'') would help define new domains of risk in a population. Discovering patient subgroups with the similar underlying disease patterns could facilitate diagnosis and treatment decision making, and epidemiological analysis and research~\cite{schnell2016bayesian}.

In this study, we utilized Latent Dirichlet Allocation (LDA), an unsupervised generative probabilistic models, and proposed a novel model named Poisson Dirichlet Model (PDM), which extends the LDA approach for the EHR data. The proposed PDM uses a Poisson distribution to model patients' disease diagnoses by considering both observed and expected observations that alleviates the impact of age and sex on a population. In the experiments, we evaluated LDA and PDM on the EHR data of three patient cohorts, namely Osteoporosis, Delirium/Dementia, and Chronic Obstructive Pulmonary Disease (COPD)/Bronchiectasis Cohorts, retrieved from the Rochester Epidemiology Project (REP) medical records linkage system, for the discovery of latent disease clusters and patient subgroups. We compared the effectiveness of LDA and PDM in identifying latent disease clusters through the visualization of disease representations learned by two approaches. We also tested the performance of LDA and PDM in differentiating patient subgroups through survival analysis, as well as statistical analysis of demographics and Elixhauser Comorbidity Index (ECI) scores in those subgroups. The experimental results show that the proposed PDM could effectively identify distinguished disease clusters based on the latent patterns hidden in the EHR data by alleviating the impact of age and sex, and that LDA could stratify patients into more differentiable subgroups than PDM in terms of p-values. However, the subgroups discovered by PDM might imply the underlying patterns of diseases of greater interest in epidemiology research due to the alleviation of age and sex. Therefore, both unsupervised machine learning approaches could be leveraged to discover patient subgroups using EHRs but with different foci.

\section{Background}

Many integrative studies have assessed the relation and impact of multimorbidities, which could reveal disease clusters, potential pathobiology mechanisms, and improve our understanding of conditions in older people~\cite{divo2014ageing}. For example, Barab\'asi et al. used the mathematical network theory to demonstrate disease coexistence in a graph model~\cite{barabasi2011network}. However, the disease network relies on the biological networks and interaction resources, and is limited due to the incompletion of our genome and phoneme knowledge ~\cite{ni2018constructing}. The wide adoption of EHR data has provided an unprecedented opportunity to apply data-driven analysis to uncover new links in diseases and patients~\cite{gligorijevic2016large}. However, the complete capture of all diagnostic events in a geographically-defined population is a key factor for discovering disease patterns. 

The creation of medical record linkage systems that connect EHR data from multiple institutions could capture the entire health care experience of a geographically defined population. The REP is a pioneer linkage system developed through a collaboration between health care providers in southeastern Minnesota, and involves Olmsted Medical Center, Mayo Clinic, Rochester Family Medicine Clinic and other medical care providers in southeastern Minnesota~\cite{melton1996history, rocca2012history, st2012data}. It is a unique infrastructure for epidemiology and outcomes research that links the medical records of local health care providers to community residents. Enabled by the REP, previously studies evaluated morbidity occurrence one diagnosis at a time~\cite{melton2013long} and used traditional analytic techniques (e.g., tree models/recursive partitioning) to define disease clusters~\cite{savica2013risk}. However, these approaches do not adequately address the co-occurrence of multiple disease states within an individual. 

A set of methods relevant to studying multimorbidities has arisen in the field of document processing under the rubric of ``topic models''. Latent Dirichlet Allocation (LDA) is a typical topic modeling method proposed by Blei et al.~\cite{blei2012probabilistic}. LDA categorizes all words in a collection of textual documents into a set of distinct ``topics'', while simultaneously classifying each document by the topics it contains~\cite{zhao2014topic}. A given word may be associated with multiple topics, and multiple topics may appear in a given document. Since a patient could also be represented by a set of diagnosed diseases that share similar undiscovered interrelations, we simply used the analogy ``words''$=$``diseases'', ``documents''$=$``patients'', ``topics''$=$``latent disease clusters'', and applied LDA to discover latent disease clusters in our previous work~\cite{li2014discovering}. Although our previous study showed the potential of LDA in leveraging hidden pattern information from EHR data, we failed to observe dichotomized latent disease clusters from the results of LDA. Moreover, another shortcoming of LDA, especially in a cohort that spans a large age range, is that it will identify clusters that are due primarily to age and/or sex, disease associations that are already known and thus not very interesting. Examples would be athletic injuries or vaccinations in the young and joint ailments in the old. LDA uses a Multinomial distribution to simulate the generative process of each single disease, which leads LDA to focus on the proportions of various diseases in the cohort.  Of greater interest in epidemiology research is the prediction or clustering of \textit{excess} risk, event rates that are over and above what would be expected for a given age and sex. Furthermore, we didn't investigate the potential of topic models in discovering patient subgroups for a defined cohort. Stratifying patients into subgroups with similar characteristics and risks will not only facilitate epidemiological analysis and research, but enable personalized care that will improve the efficiency and effectiveness of disease prevention, diagnosis, and treatment.

\section{Methods}

\subsection{Mathematical Modeling}


Suppose $D$ denotes a disease diagnosis code set $D = \{d_1,d_2,...,d_V\}$ with size $V$, a cohort $C$ is represented as a group of $M$ patients $C=\{\mathbf{w}_1,\mathbf{w}_2,...,\mathbf{w}_M\}$, and a patient $\mathbf{w}_m$ is represented as a sequence of $N$ disease diagnoses $\mathbf{w}_m=(w_{m,1},w_{m,2},...,w_{m,N})$ where $w_{m,n} \in D, n=1,2,...,N$ is a disease diagnosis code from $D$ for patient $\mathbf{w}_m$. Given these notations, we describe LDA and the proposed PDM in this subsection.



Let $z$ denote the latent disease clusters, which is akin to topics in LDA. Suppose $K$ denotes the dimensionality of $z$, $\theta$ a $K$-dimensional Dirichlet random variable, $\alpha$  a $K$-dimensional parameter with $\alpha_i>0$, and $\beta$  a $K\times V$ matrix parameter, we can define LDA as the following generative process for each patient $\mathbf{w}$ in a cohort $C$:
\begin{enumerate}
  \item For each of $K$ latent disease clusters:
    \begin{enumerate}
      \item Choose $\phi_k$ $\sim$ Dirichlet$(\beta)$
    \end{enumerate}
  \item For each of $M$ patients in $C$:
    \begin{enumerate}
      \item Choose $\theta_m$ $\sim$ Dirichlet($\alpha$).
      \item For each of $N$ diseases that patient $\mathbf{w}_m$ has been diagnosed:
        \begin{enumerate}
	      \item Choose a latent disease cluster $z_{m,n}$ $\sim$ Multinomial($\theta_m$).
	      \item Choose a disease $w_{m,n}$ $\sim$ Multinomial($\phi_{z_{m,n}}$).
        \end{enumerate}
   \end{enumerate}
\end{enumerate}

\begin{figure}
\centering
\includegraphics[width=\textwidth]{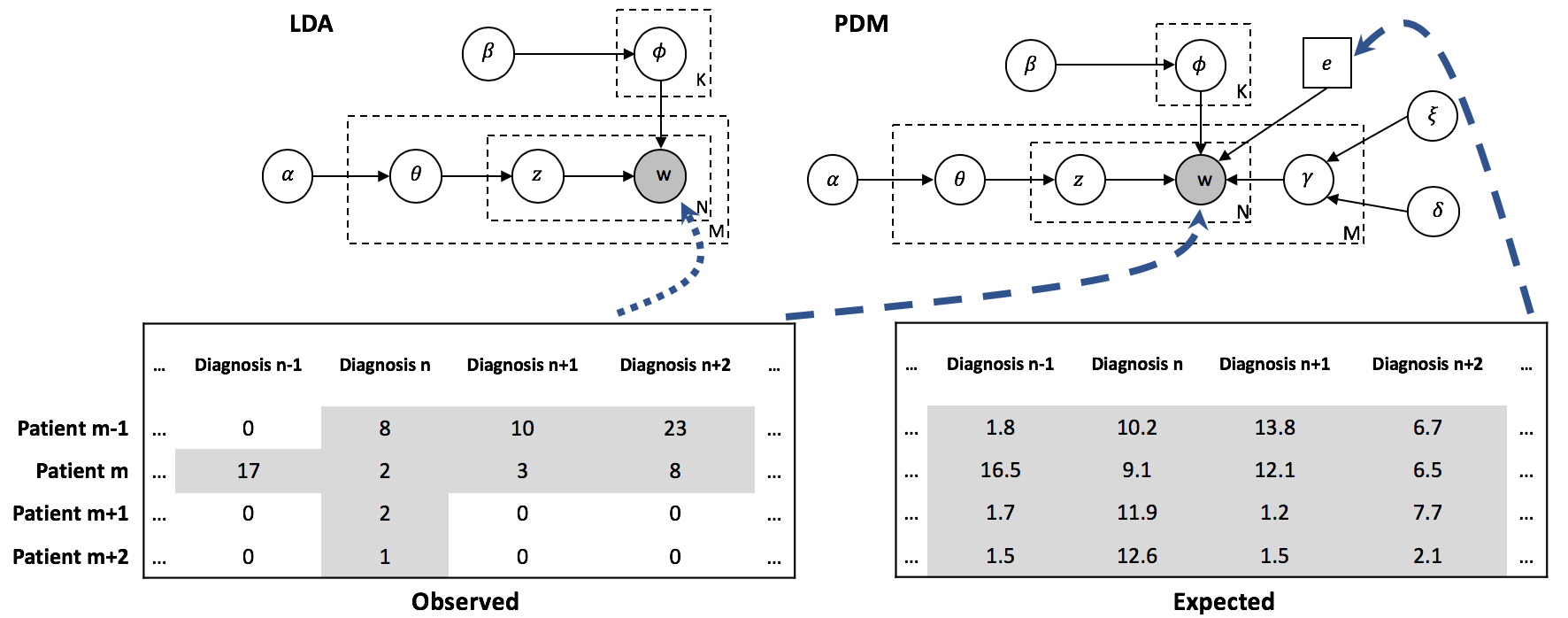}
\caption{The graphical model representation of LDA (upper-left) and PDM (upper-right). Circles represents random variables, gray-shaded circles represents observed states, the dashed plates represents replication over a set of variables, and the solid squares represent constants, which specify fixed-valued variables. $M$ is the number of patients in a cohort, $N$ is number of diagnosis for a patient, and $K$ is the number of disease groups. The proposed PDM leverages both observed and expected number of diagnosis for each patient in the population, which alleviates the difficulty of LDA in dealing with missing data that were not collected during residents' absence in the healthcare system, particularly in the medical EHRs, as well as incorporating the epidemiological characteristics of the population. }
\label{fig.model}
\end{figure}

LDA can be represented as a graphical model at the upper-left of Figure \ref{fig.model}. By applying Gibbs sampling, we can learn parameters $\alpha$ and $\beta$, and hyper-parameters $\theta$ and $\phi$ in LDA.  We could leverage $\phi$, the probability of a disease in a latent topic, to discover latent comorbidities that appear in the same disease cluster. 




In this study, we also propose a novel unsupervised machine learning model, Poisson Dirichlet Model (PDM), which extends LDA for discovering disease clusters of excess risk. Based on a patient's age, sex, and length of follow-ups we can compute an expected number of diagnoses $e_{m,n}$ for subject m and diagnosis n, and compare the observed count $y_{m,n}$ to this expectation. We hypothesize that the PDM could be sensitive to patterns of excess disease risk, which will be different than overall risk, and that these patterns could identify more distinguished disease clusters than LDA.  

With the same notations used for LDA, PDM assumes the following generative process for each patient in the cohort:
\begin{enumerate}
  \item For each of $K$ latent disease clusters:
    \begin{enumerate}
      \item Choose $\phi$ $\sim$ Dirichlet$(\beta)$
    \end{enumerate}
  \item For each of $M$ patients in $C$:
    \begin{enumerate}
      \item Choose $\gamma_m$ $\sim$ Gamma($\xi$, $\delta$), where $E(\gamma)=\xi\cdot \delta=1$.
      \item Choose $\theta_m$ $\sim$ Dirichlet($\alpha$).
      \item For each of $N$ diseases that patient $\mathbf{w}_m$ has been diagnosed:
        \begin{enumerate}
	      \item Choose a latent disease cluster $z_{m,n}$ $\sim$ Multinomial($\theta_m$).
	      \item Choose a disease count $y_{m,n}$ $\sim$ Poisson($\phi_{z_{m,n}} \cdot  e_{m,n} \cdot \gamma_m$).
        \end{enumerate}
   \end{enumerate}
\end{enumerate}

The proposed PDM model is represented as a probabilistic graphical model at the upper-right of Figure \ref{fig.model}. In this generative process, $e_{m,n}$ denotes the expected number of diagnoses for disease $w_{m,n}$ that is computed by a simple rate model fit to the overall data. Specifically, the follow-ups for each subject in the cohort were divided into bins based on single years of age and sex. For each disease diagnosis, the counts (number of occurrences) were modeled using a generalized additive model separately for males and females, assuming a Poisson error structure.  Age was fit using a smoothing spline with 4 degrees of freedom and the log of person-years in each bin were treated as an offset.  The expected event rates for each person were estimated using predictions from these models based on the age, sex, and follow-up of each person. After dividing each patient's follow-up by age and sex, $e_{m,n}$ can be derived mathematically by:
$$e_{m,n} = \mathbb{E}(Y|X_{w_{m,n}})=\beta + log(X_{1,{w_{m,n}}}) + X_{2,{w_{m,n}}}$$
where $\beta$ is a coefficient, $log(X_{1,{w_{m,n}}})$ is an offset of the person-years,  $X_{2,{w_{m,n}}}$ is age, and $Y$ follows a Poisson distribution. 

Poisson distribution is utilized to generate the total number of each disease that has been diagnosed for a patient. $\gamma_m$ is a positive multiplier for patient $m$ generated from a Gamma distribution with mean equal to $1$. Since some patients may have a significantly larger number of diagnosis (e.g., sicker or better insurance) than others, $\gamma_m$ is utilized as a normalizer on the number of patients' diagnosis so that the parameters of Poisson distribution are not learned towards the extreme cases.  

Figure \ref{fig.model} also illustrates that LDA models the observed number of disease diagnoses while PDM takes advantage of the combination of observed and expected number of diagnoses. The use of expected data for each patient in PDM alleviates the impact of age and sex, and lessen the difficulty of LDA in handling missing data that were not collected during residents' absence in the healthcare system, particularly in medical EHRs, while simultaneously incorporates the epidemiological characteristics of the population. Since the experimental data are from Olmsted County, we used the expected risk table of the Minnesota population. Since $e_{m,n}$ is pre-calculated before training PDM, it is treated as a fixed-valued constant in the parameter learning process.

\subsection{Parameter Estimation}

Markov Chain Monte Carlo (MCMC) methods are usually used to generate random samples that can be used in estimating the parameters of posterior distributions in a probabilistic machine learning model. MCMC constructs a Markov chain to converge to the target distribution, and generates samples from that Markov chain. Since the Dirichlet priors are conjugate to the Multinomial distributions, Gibbs sampling, a widely adopted MCMC algorithm, is utilized for inference of the LDA model~\cite{griffiths2004finding}. However, it cannot be applied to PDM since the Poisson distributions are not conjugate to the Dirichlet priors. Therefore, a more general MCMC sampling method, Metropolis-Hastings (MH) algorithm, was applied to approximate the distributions and learn the parameters of PDM. The MH sampling algorithm creates a Markov chain based on a proposal distribution and corrects the wrong density through an acceptance-rejection step, in comparison with the Gibbs sampling algorithm that always accepts the proposal distribution. We briefly introduce the MH algorithm as below.

Suppose that $P(x)$ denotes the true distributions we would like to inference, $Q(x)$ denotes the proposal distribution, $x^*$ is a new sample drawn from $Q(x^*|x)$, and $A(x^*|x)$ is the acceptance probability for $x^*$ defined by 
$$A(x^*|x)=\min{ \left(1,\frac{P(x^*Q(x|x^*))}{P(x)Q(x^*|x)}\right)},$$
the MH sampling algorithm is illustrated in Algorithm \ref{alg.1}. We implemented the parameter estimation algorithm using JAGS\footnote{http://mcmc-jags.sourceforge.net/} and rJAGS\footnote{https://cran.r-project.org/package=rjags}, which automatically choose the  proposal distribution for the sampling process. We utilized two sampling chains with a burn-in of 500 iterations followed by 1000 iterations for inference. 

\begin{algorithm}
  \caption{The Metropolis-Hastings Algorithm}
  \label{alg.1}
  \begin{algorithmic}[1]
  \STATE Initialize starting state of a Markov chain at $x^0$, set $t=0$.
  \STATE $x=x^t$
  \STATE $t=t+1$
  \STATE sample $x^*$ from the proposal distribution, i.e., $x^*$ $\sim$ $Q(x^*|x)$
  \STATE sample $u$ from a Uniform distribution, i.e., $u$ $\sim$ Uniform($0,1$)
  \IF{$u\leq A(x^*|x)$}
        \STATE{$x^t=x^*$}
  \ELSE 
    \STATE{$x^t=x$}
  \ENDIF
  \end{algorithmic}
\end{algorithm}

\subsection{Applications on EHRs}

\subsubsection{Discovering Latent Disease Clusters}

Given a cohort of patients diagnosed with a certain disease, unsupervised machine learning approaches allow us to discover latent comorbidity clusters for that disease, which could help define new domains of risk. LDA and the proposed PDM model represent diseases in a latent topic space. The estimated parameter $\beta$ in LDA or PDM is a disease-topic matrix indicating the probability of a disease occurring in a latent topic. We hypothesize that the diseases with similar characteristics would be automatically clustered in the latent topic space, which is called a latent disease cluster. In order to verify the effectiveness of LDA and PDM for identifying latent disease clusters, we qualitatively visualize the disease representation in the latent topic space using the disease-topic matrix by applying a machine learning visualization method, t-SNE~\cite{maaten2008visualizing}, which maps the disease representation in the high-dimensional topic space into a two-dimensional space that enables visualization. By doing so, we intuitively evaluate the potential of unsupervised machine learning in discovering disease clusters. Since the perplexity and the number of iterations are two important parameters for the t-SNE, we tested different combinations and chose the ones generating the best visualization results for PDM and LDA.

\subsubsection{Discovering Patient Subgroups}

An interesting question in epidemiology study is whether we can stratify patients into subgroups so that the patients in the same subgroup have similar health characteristics and risks. Population-based evidence has been shown to be a major source of support for medical decision making for an individual~\cite{ledbetter2001toward}. Using the estimated parameters of LDA or PDM, we can calculate the posterior probability of a latent disease cluster for a given patient, i.e., $p(z|\mathbf{w}_m, \alpha, \beta)=\sum_i^N p(z|w_{m,i}, \alpha, \beta)$, which is a patient-topic probability matrix. In order to discover patient subgroups, we could leverage clustering analysis on the patient feature vectors by using the rows of patient-disease cluster matrix. In our experiments, we tested three clustering algorithms, namely hierarchical clustering~\cite{ward1963hierarchical}, K-means clustering~\cite{hartigan1979algorithm}, and Birch clustering algorithms~\cite{zhang1996birch}, with five different numbers of subgroups, ranging from 2 to 6 subgroups. 

To evaluate these subgroups, we carried out survival analysis on each patient subgroup. We used the Log-rank statistical test to compare the difference between the survival curves of patient subgroups. In addition to the survival analysis, we conducted statistical analysis on the demographics and number of diagnoses for patient subgroups. We also used the widely adopted comorbidity measure, Elixhauser Comorbidity Index (ECI)~\cite{elixhauser1998comorbidity}, to compare patient subgroups in each cohort. We compared ECI scores between patient subgroups, and 29 ECI categories, including congestive heart failure, cardiac arrhythmias, valvular disease, pulmonary circulation disorders, peripheral vascular disease, hypertension, paralysis, other neurological disorders, chronic pulmonary disease, diabetes (uncomplicated), diabetes (complicated), hypothyroidism, renal failure, liver disease,  peptic ulcer disease (excluding bleeding), lymphoma, metastatic cancer,  solid tumour (without metastasis),  rheumatoid arthritis/collaged vascular disease, coagulopathy, obesity, weight loss, fluid and electrolyte disorders, blood loss anaemia,  deficiency anaemia, alcohol abuse, drug abuse, psychoses, and depression. Our goal is to evaluate whether the patient subgroups discovered by the proposed PDM model could differentiate patients in a defined cohort.

\section{Datasets}

In this section, we describe the datasets used in the empirical experiments of testing LDA and the proposed PDM approach. Three cohorts extracted from the REP, namely the Osteoporosis Cohort, the Delirium/Dementia Cohort, and the Chronic Obstructive Pulmonary Disease (COPD)/Bronchiectasis Cohort, were utilized to evaluate the proposed PDM model. These cohorts were retrieved from the REP cohort that consisted of patients 50 years of age and older during the interval January 1, 1995 through December 31, 2011 from Olmsted County whose total number of disease diagnoses was between three hundred and five hundred and who had at least thirty diagnoses of osteoporosis (CCS code: 206), delirium/dementia (CCS code: 653), and COPD/bronchiectasis (CCS code: 127), respectively. These diseases have been shown to be related to aging with complicated comorbidities. We used Clinical Classifications Software (CCS)\footnote{https://www.hcup-us.ahrq.gov/toolssoftware/ccs/ccs.jsp} as the diagnosis taxonomy for each patient since the International Classification of Diseases (ICD) classification has too detailed granularity for clinical practice ~\cite{li2014discovering}. Since the REP had created a longitudinal record spanning each subject's entire period of residency in the community, we retrieved all their ICD-9 diagnostic codes from the REP linkage system. We collapsed the ICD-9 codes into 285 CCS categories and removed the less frequent codes. Table \ref{tab.dem} lists the basic demographics of the three cohorts, including the number of patients (male and female), median age (male and female), and median observed number of diagnoses for the three cohorts. LDA and PDM were trained on the data of these cohorts, with five different numbers of latent disease clusters ($K=10,20,30,40,50$). Eventually $K=20$ was chosen for both approaches based on outperforming experimental results.

\begin{table}[!h]
    \centering
    \begin{tabular}{l|cccc}
    \hline
    Cohort & Osteoporosis Cohort & Delirium/Dementia Cohort & COPD/Bronchiectasis Cohort  \\
    \hline
    \# Patients & 388 & 304 & 685  \\
    \# Male (\%) & 21 (5.4\%) & 95 (31.2\%) & 337 (49.2\%)  \\
    \# Female (\%) & 367 (94.6\%)& 209 (68.8\%) & 348 (50.8\%)  \\
    Median Age & 74.4 & 83.6 & 73.2  \\
    Median Age (Male) & 74.7 & 85.0 & 75.1  \\
    Median Age (Female) & 68.8 & 81.6 & 71.1  \\
    Median Observed \# Diagnosis & 406 & 387.5 & 402  \\
    \hline
    \end{tabular}
    \caption{Demographics and basic statistics of the study cohorts.}
    \label{tab.dem}
\end{table}

\section{Results}

\subsection{Visualization of Latent Disease Clusters}

\begin{figure}
\centering
  \begin{tabular}{p{1.5cm}cc}
   & LDA & PDM  \\
Osteoporosis Cohort& \raisebox{-.5\height}{\includegraphics[width=0.3\textwidth]{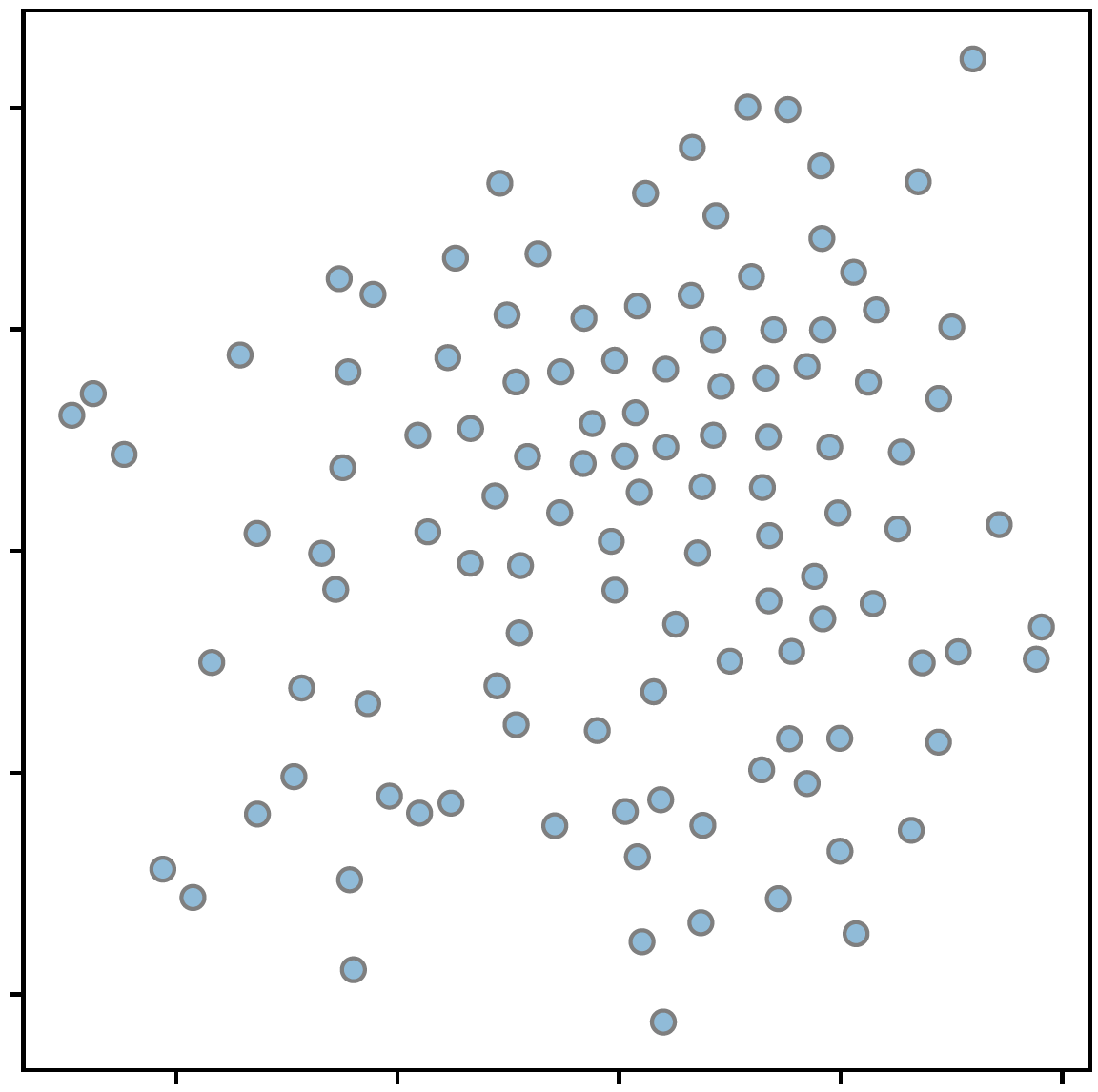}} &\raisebox{-.5\height}{\includegraphics[width=0.3\textwidth]{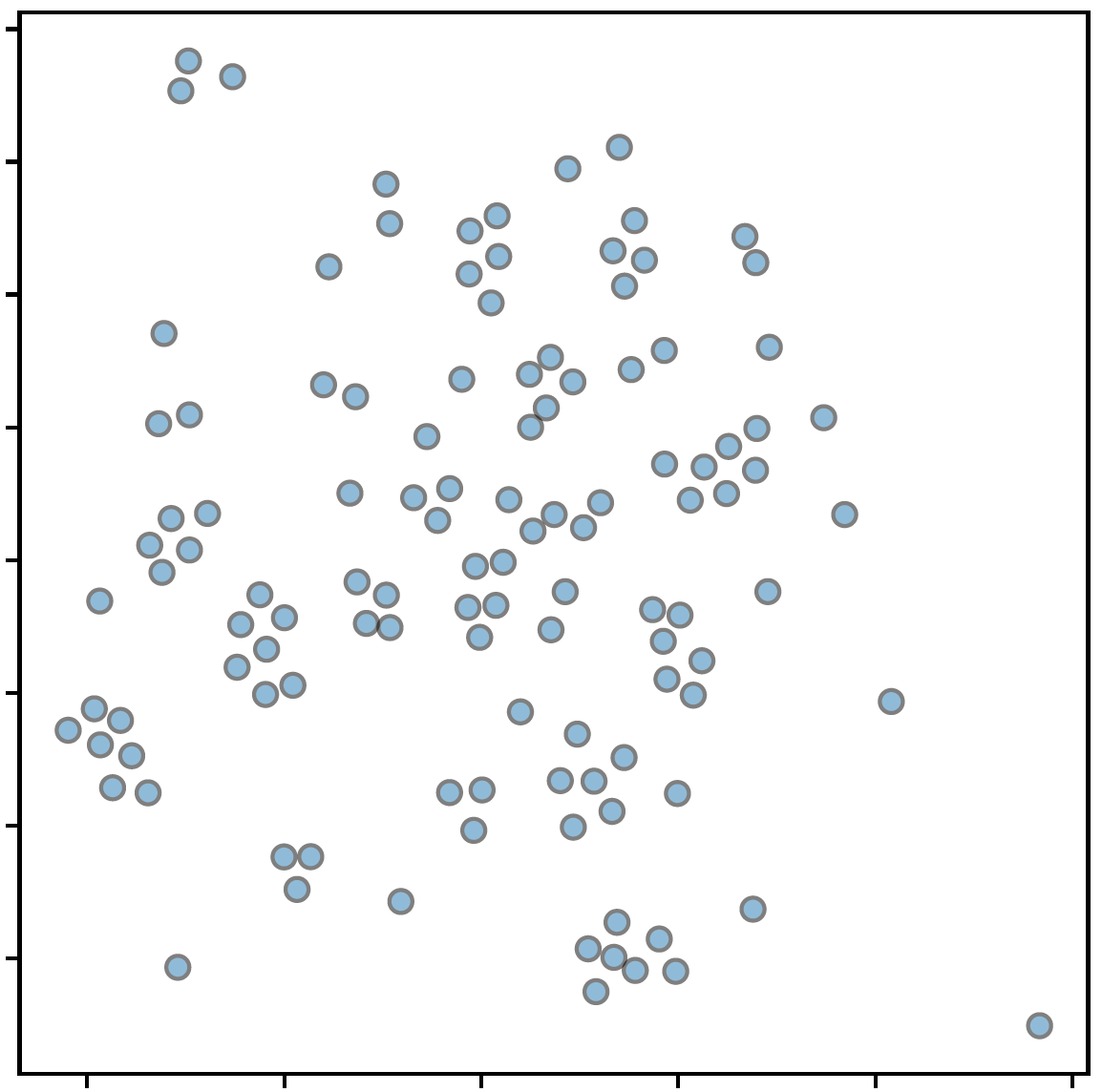}} \\
Delirium/Dementia Cohort& \raisebox{-.5\height}{\includegraphics[width=0.3\textwidth]{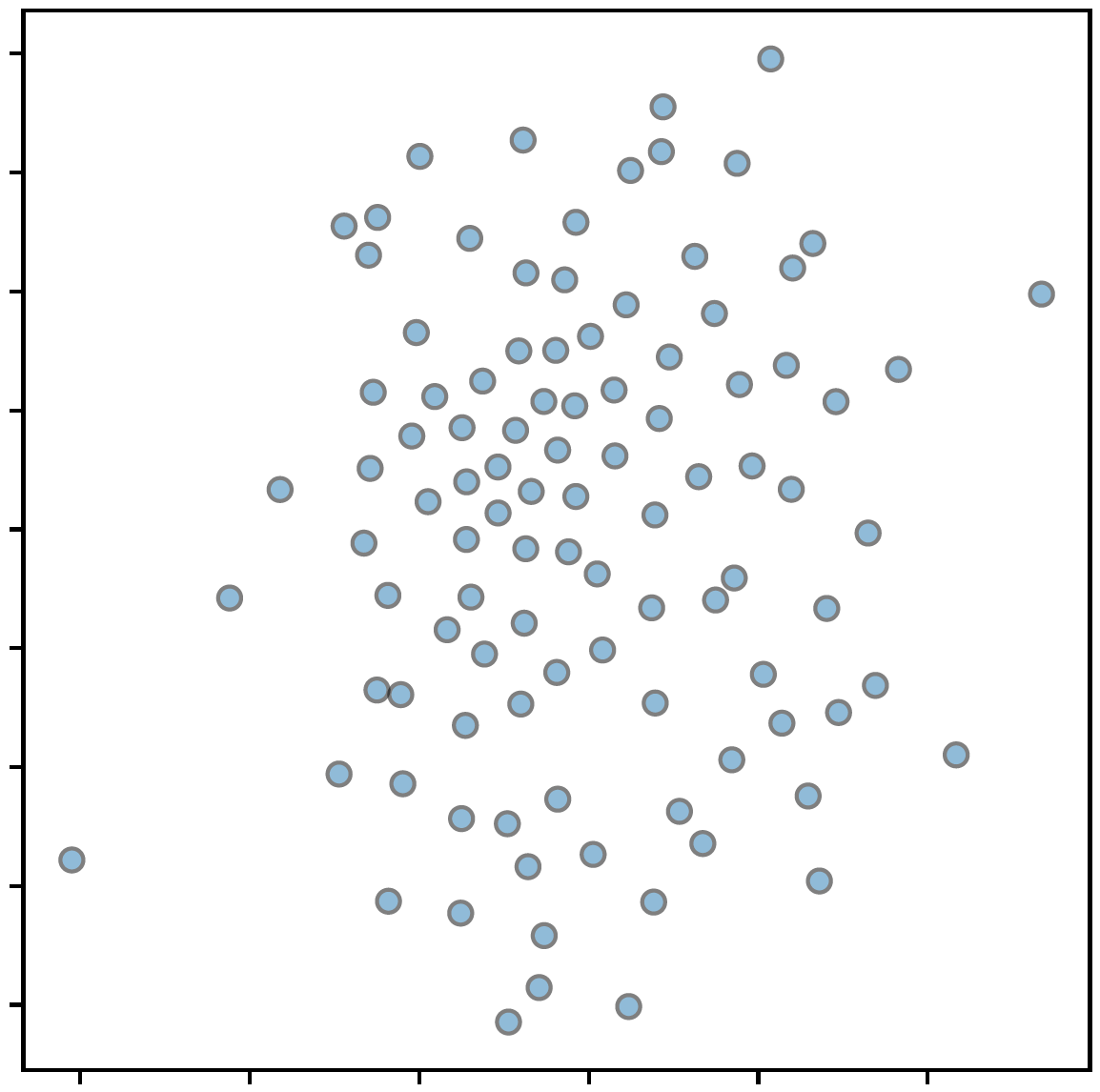}} &\raisebox{-.5\height}{\includegraphics[width=0.3\textwidth]{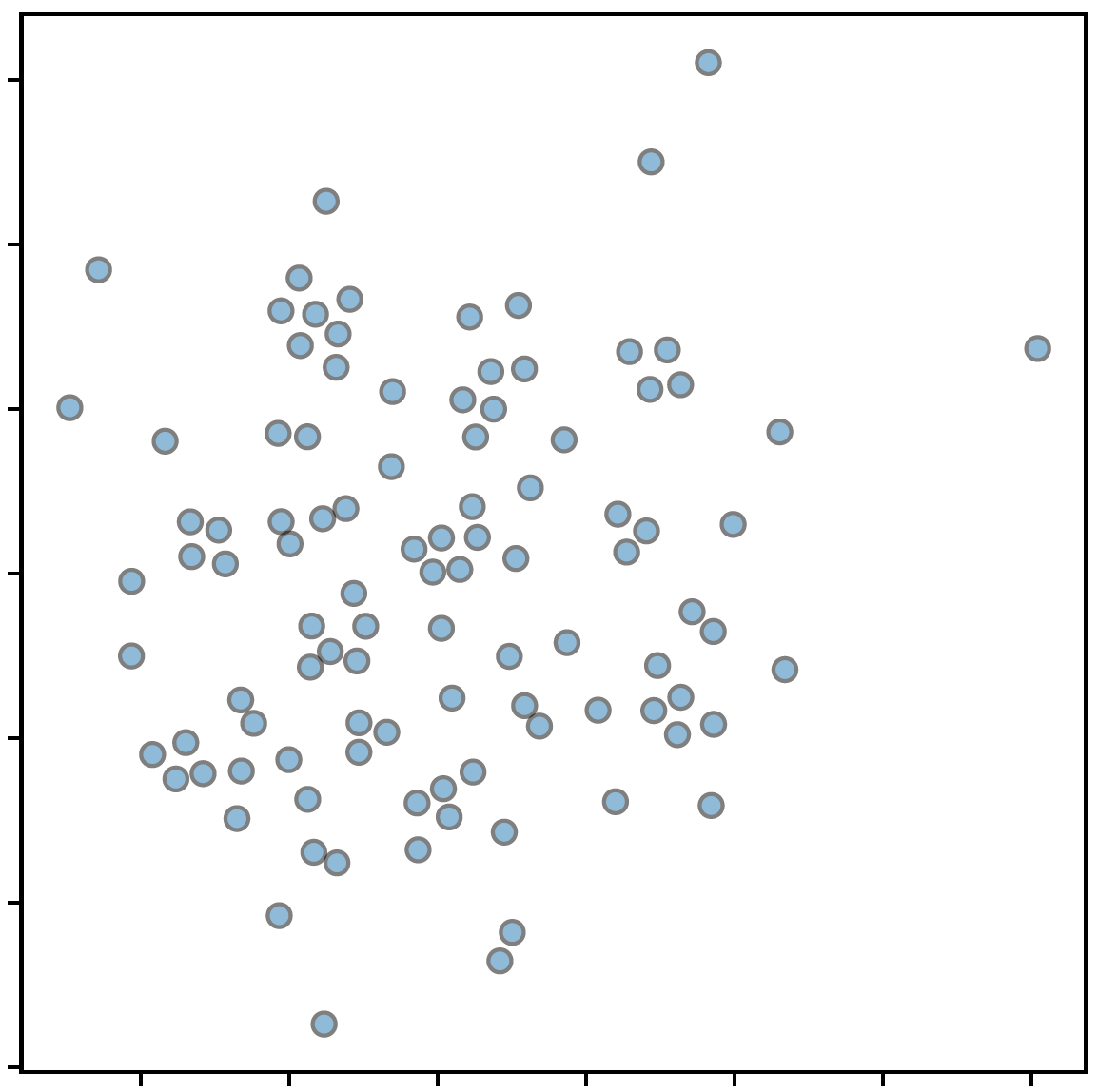}}  \\
COPD/Bronchiectasis Cohort& \raisebox{-.5\height}{\includegraphics[width=0.3\textwidth]{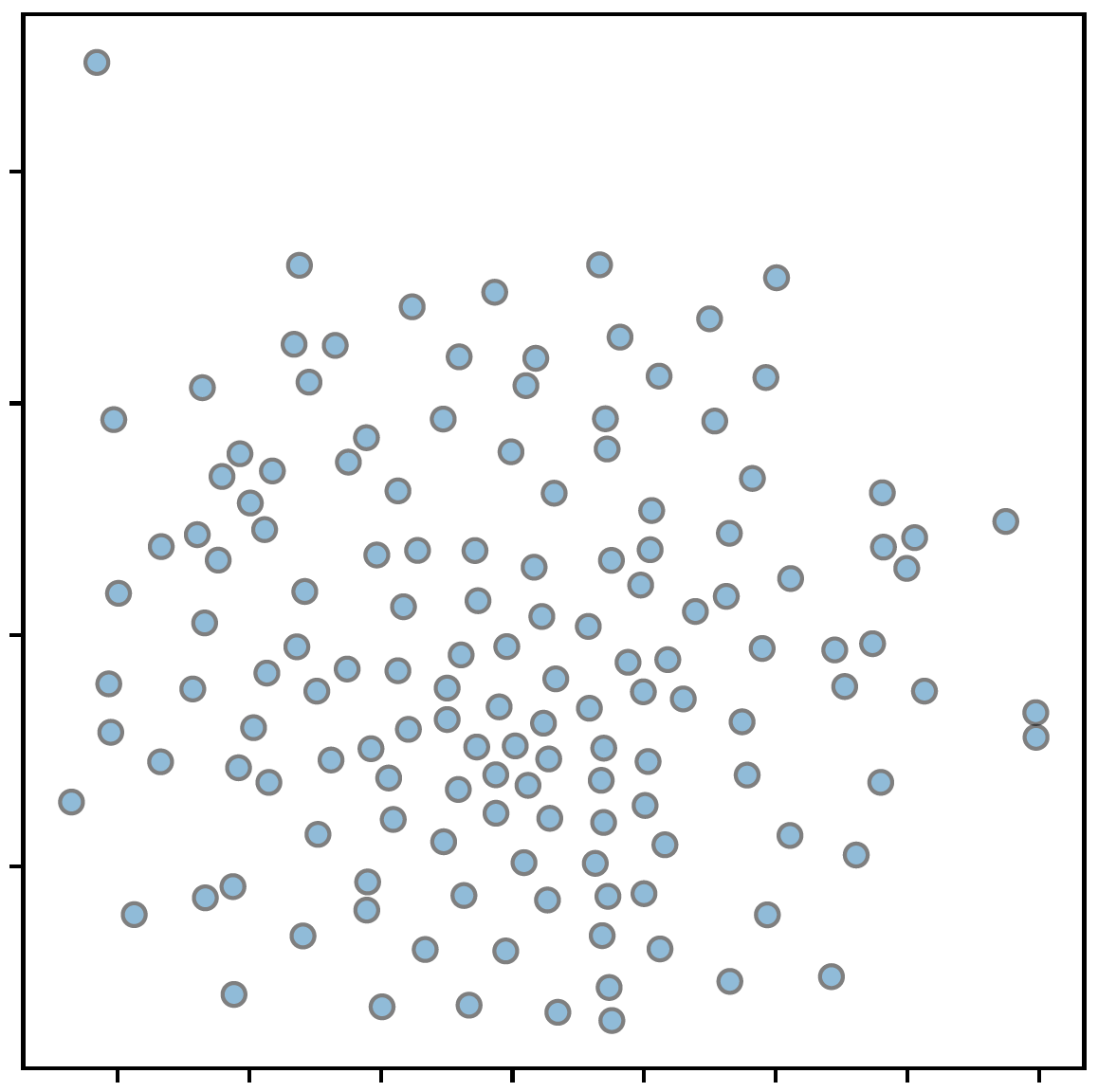}} &\raisebox{-.5\height}{\includegraphics[width=0.3\textwidth]{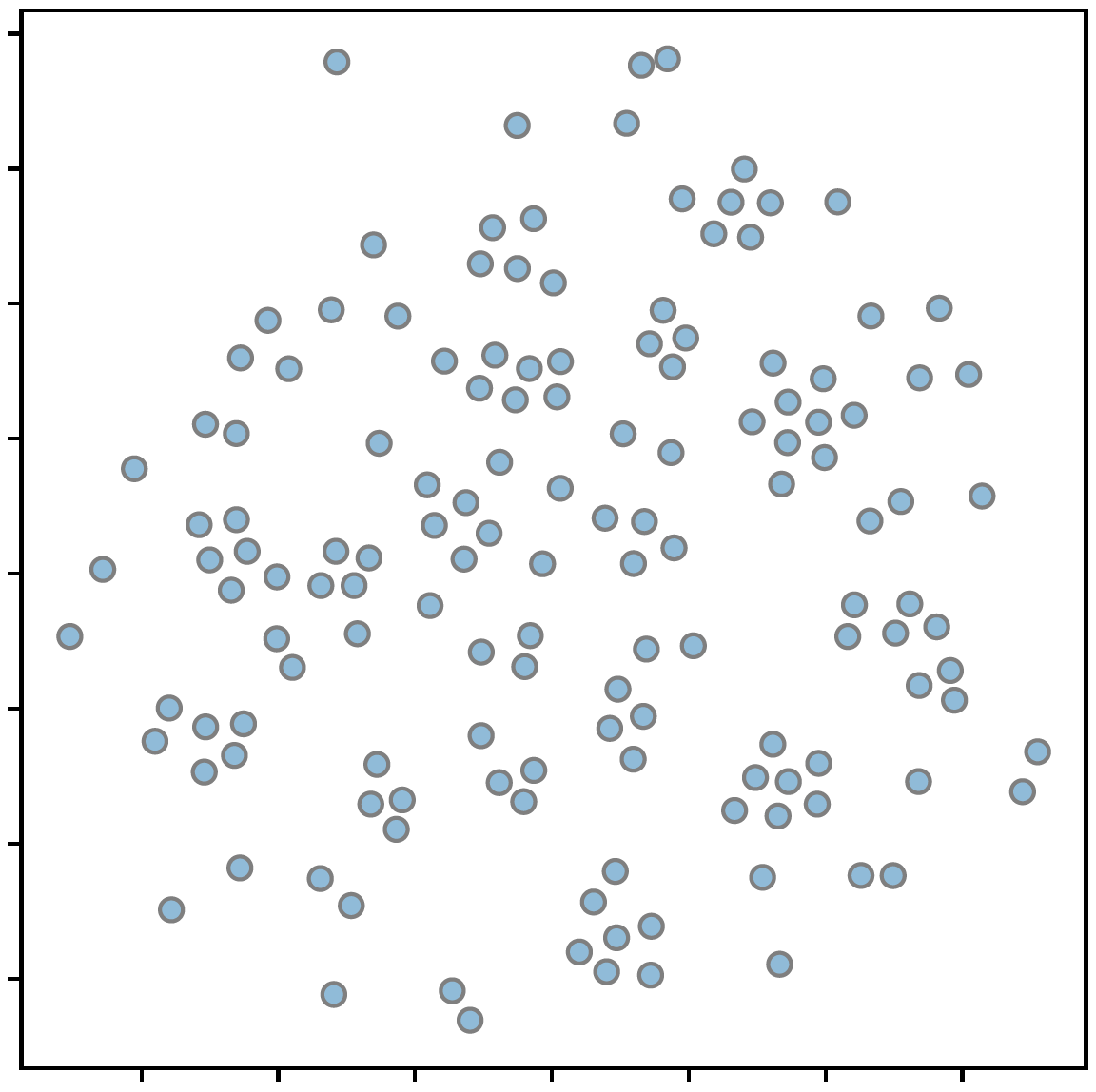}}  \\
  \end{tabular}
  \caption{Comparison of the visualization of diseases (represented by CCS) in the latent topic space learned by LDA (left column) and PDM (right column) for three cohorts. }
  \label{fig.tsne}
\end{figure}

We first compared the visualization of disease representations in the latent topic space learned by LDA and the proposed PDM model for three cohorts, as depicted in Figure \ref{fig.tsne}.  The perplexity parameters of t-SNE for PDM and LDA are 10 and 20, respectively, and the number of iterations is 5000 for both approaches. It is obviously shown that the disease clusters are explicitly dichotomized by PDM while almost no disease clusters could be observed by LDA. Since LDA failed to identify latent disease clusters, we list the latent comorbidities that are clustered with osteoporosis, delirium/dementia, and COPD by PDM in Table \ref{tab.comorb}. In order to verify the results, we tried to find evidence by searching PubMed\footnote{https://www.ncbi.nlm.nih.gov/pubmed/} articles' title or abstract using keywords from the target disease and the latent comorbidities, in combination with the term ``comorbidity'' or ``comorbidities''. For example, we found 31 PubMed research articles for \textit{osteoporosis} and \textit{implant or graft}, 52 articles for  \textit{dementia} and  \textit{osteoporosis}, and 30 articles for  \textit{COPD} and  \textit{cerebrovascular disease}. The latent comorbidities are not highly related to age and sex and thus the prediction or clustering of \textit{excess} risk would be of more interest for epidemiological analysis. This result shows that the proposed PDM model is able to learn the latent patterns hidden in the EHR data that differentiate disease clusters by alleviating the impact of age and sex on the diseases. 

\begin{table}
    \centering
    \begin{tabular}{l|p{0.5\textwidth}}
    \hline
    Disease & Latent Comorbidities \\
     \hline
    Osteoporosis (CCS: 206) & Headache; including migraine (CCS: 84) \newline
Nonspecific chest pain (CCS: 102) \newline
Diverticulosis and diverticulitis (CCS: 146) \newline
Complication of device; implant or graft (CCS: 237) \newline
Abdominal pain (CCS: 251) \\
\hline
Delirium/dementia (CCS: 653) & Immunizations and screening for infectious disease (CCS: 10) \newline
Conduction disorders (CCS: 105) \newline
Osteoporosis (CCS: 206) \newline
Fracture of lower limb (CCS: 230) \\
\hline
COPD/bronchiectasis (CCS: 127) & Peri-; endo-; and myocarditis; cardiomyopathy (except that caused by tuberculosis or sexually transmitted disease) (CCS:97) \newline
 Aortic; peripheral; and visceral artery aneurysms (CCS: 115) \newline
 Fracture of lower limb (CCS: 230) \newline
 Late effects of cerebrovascular disease (CCS: 113) \newline
 Complications of surgical procedures or medical care (CCS: 238) \newline
 Biliary tract disease (CCS: 149) \newline
 Other female genital disorders (CCS: 175) \\
    \hline
    \end{tabular}
    \caption{Latent comorbidities of osteoporosis, delirium/dementia, and COPD/bronchiectasis discovered by the proposed PDM approach.}
    \label{tab.comorb}
\end{table}

\subsection{Validation of Patient Subgroups}

In this section, we demonstrate the experimental results of patient subgroups discovered by LDA and the proposed PDM model for three cohorts. As aforementioned, we tested three clustering algorithms (i.e., hierarchical clustering, K-means clustering, and Birch clustering algorithms) and five different numbers of patient subgroups (i.e., 2, 3, 4, 5, 6) on the patient-topic matrix computed by LDA and PDM. We carried out survival analysis on these subgroups for each cohort. The p-values of the Log-rank test on the survival curves are listed in Table \ref{tab.pvalue}. From the table, we observe that LDA generallly produces patient subgroups with smaller p-values than PDM. When LDA was utilized, the K-means clustering algorithm  generates differentiated patient subgroups with statistical significance for both Osteoporosis and COPD/Bronchiectasis Cohorts when the number of patient subgroups is 2, and achieved the best p-values ($p=0.0051$) for the Delirium/Dementia Cohort when the number of patient subgroups is 3. The p-values of patient subgroups generated by PDM were much higher than those by LDA: K-means clustering algorithm achieved p-values of 0.071 and 0.00028 with the number of patient subgroups equal to 6 and 2 for the Delirium/Dementia and COPD Cohorts, respectively; and the Birch clustering algorithm produced a p-value of 0.0085 for the Osteoporosis Cohort when the number of patient subgroups is 2. Overall, the K-means clustering algorithm outperforms other clustering algorithms in identifying patient subgroups. The patient subgroup results with the best p-values from LDA and PDM with the smallest number of patient subgroups are chosen for further survival and comorbidity analysis. 

\begin{table}
    \centering
    \small
    \begin{tabular}{l|cccccccccc}
    \hline
Number of subgroups  & \multicolumn{2}{c}{2} & \multicolumn{2}{c}{3} & \multicolumn{2}{c}{4} & \multicolumn{2}{c}{5} & \multicolumn{2}{c}{6} \\
\cmidrule(r){2-3}\cmidrule(r){4-5}\cmidrule(r){6-7} \cmidrule(r){8-9} \cmidrule(r){10-11} 
Unsupervised Methods & PDM & LDA & PDM & LDA & PDM & LDA & PDM & LDA & PDM & LDA \\
\hline
\textit{\textbf{Osteoporosis Cohort}}  \\
Hierarchical clustering  & 0.16 & 0.00017 & 0.28& 0.00081 & 0.44& 0.00075 & 0.41& $<$0.00073 & 0.41& $<$0.0001 \\
K-means clustering  & 0.39& \textbf{$<$0.0001} & 0.58& $<$0.0001 & 0.38& $<$0.0001 & 0.38& $<$0.0001 & 0.6& 0.00032 \\
Birch clustering  & \textbf{0.0085} & 0.002 & 0.015& 0.0038 & 0.0098& 0.00078 & 0.011& 0.0021 & 0.018& 0.0035 \\
    \hline
    \textit{\textbf{Delirium/Dementia Cohort}}  \\
Hierarchical clustering  & 0.10& 0.41 & 0.083& 0.23 & 0.16& 0.40 & 0.26& 0.54 & 0.28& 0.61 \\
K-means clustering  & 0.34& 0.012 & 0.49& \textbf{0.0051} & 0.50& 0.011 & 0.42& 0.029 & \textbf{0.071} & 0.0057 \\
Birch clustering  & 0.14& 0.071 & 0.34& 0.19 & 0.32& 0.34  & 0.46& 0.50 & 0.12& 0.033 \\
\hline
\textit{\textbf{COPD/Bronchiectasis Cohort}}  \\
Hierarchical clustering  & 0.017 & 0.0045 & 0.014& $<$0.0001 & 0.035& $<$0.0001 & 0.027& $<$0.0001 & 0.021& $<$0.0001 \\
K-means clustering  & \textbf{0.00028}& \textbf{$<$0.0001} & 0.00032& $<$0.0001 & 0.0017& $<$0.0001 & 0.086& $<$0.0001 & 0.0026& $<$0.0001 \\
Birch clustering & 0.15& 0.00045 & 0.10& $<$0.0001 & 0.15& $<$0.0001 & 0.12& $<$0.0001 & 0.2& $<$0.0001 \\
    \hline
    \end{tabular}
    \caption{P-values of the Log-rank test on survival curves of patient subgroups using unsupervised machine learning models and three clustering algorithms with different number of subgroups for the three cohorts. The subgroups with bolded p-values by PDM and LDA are chosen for further survival and comorbidity analysis.}
    \label{tab.pvalue}
\end{table}

\subsubsection{Survival curves}

\begin{figure}
\centering
\scriptsize
  \begin{tabular}{ccc}
    Osteoporosis Cohort & Delirium/Dementia Cohort & COPD/Bronchiectasis Cohort \\
 \raisebox{-.5\height}{\includegraphics[width=0.32\textwidth]{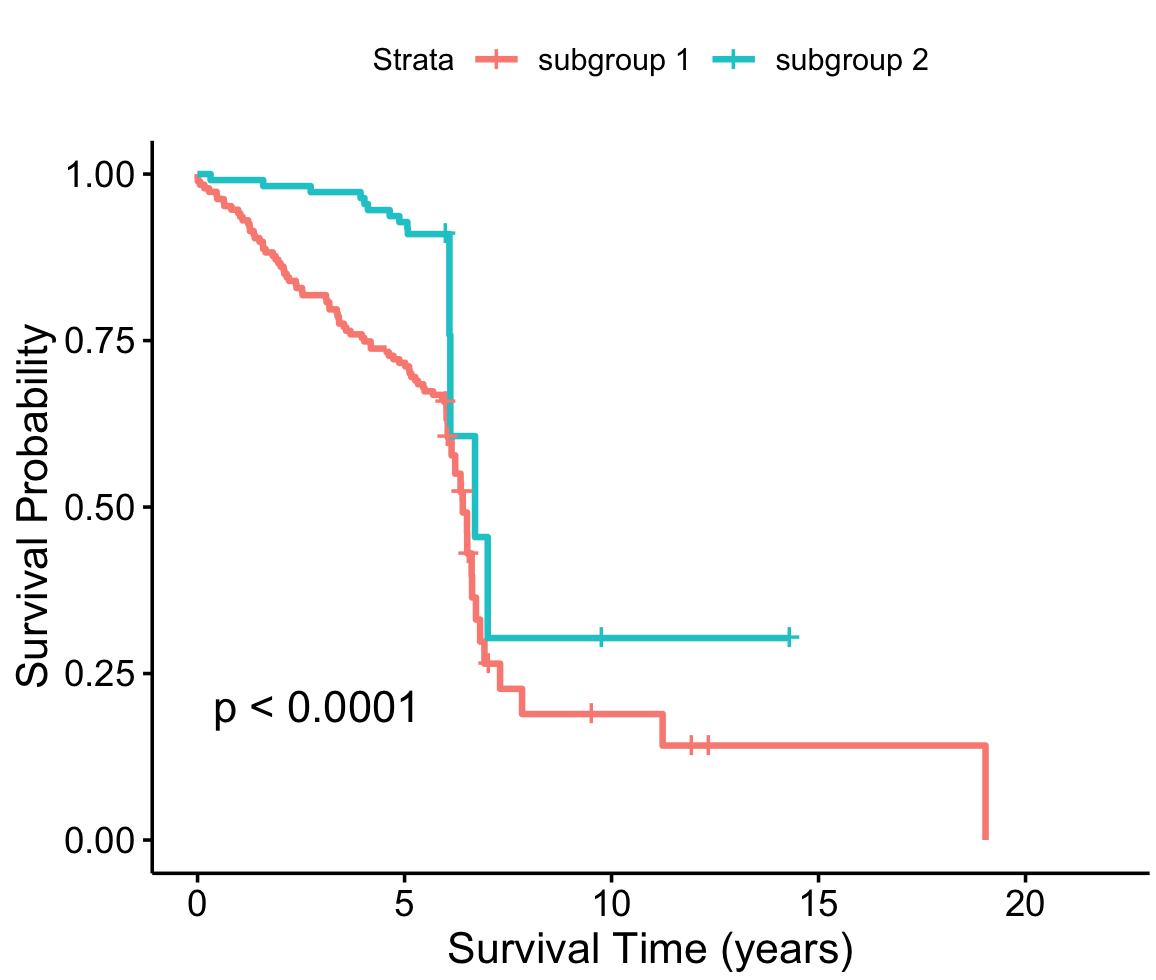}} & \raisebox{-.5\height}{\includegraphics[width=0.33\textwidth]{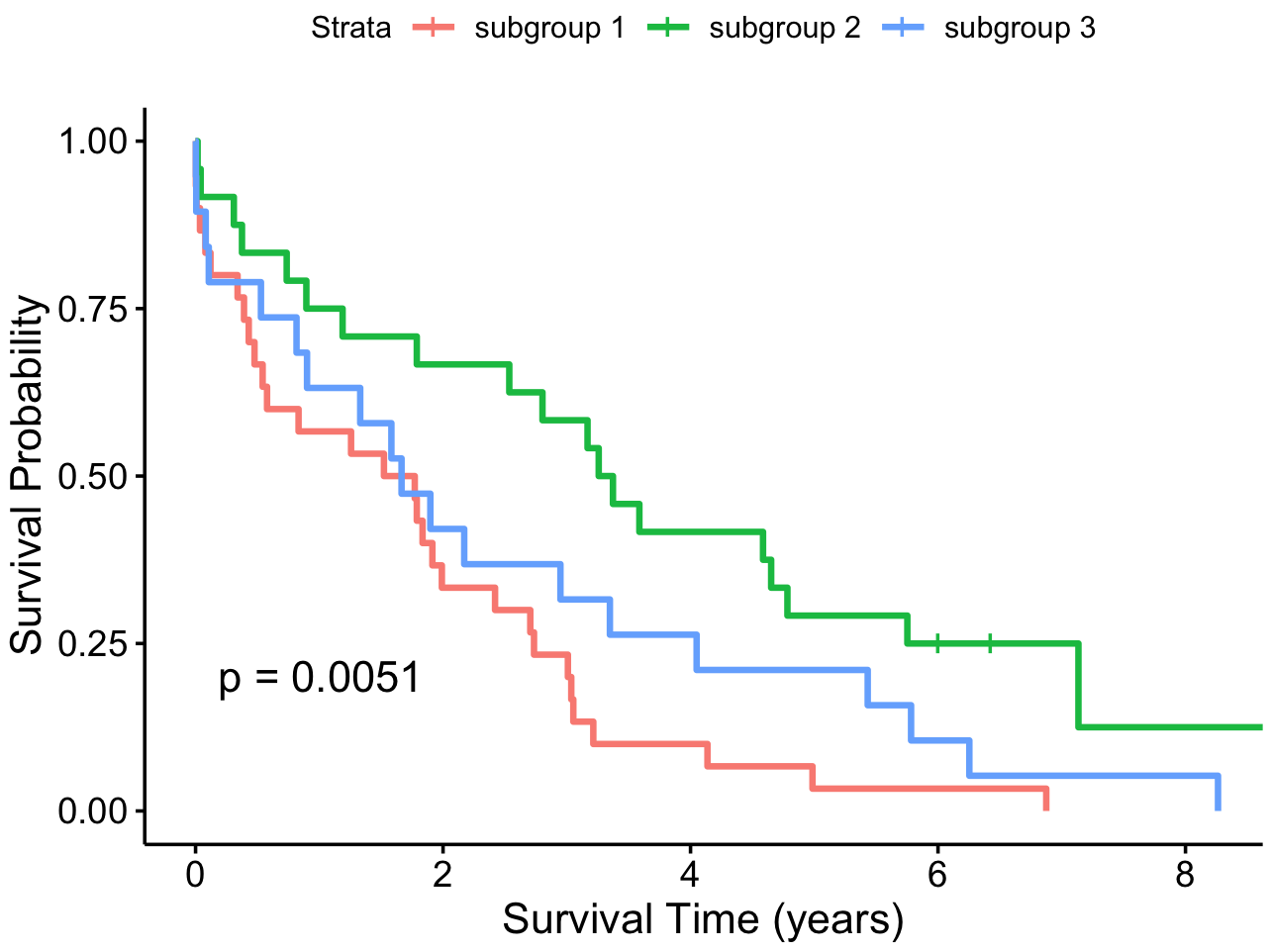}} & \raisebox{-.5\height}{\includegraphics[width=0.33\textwidth]{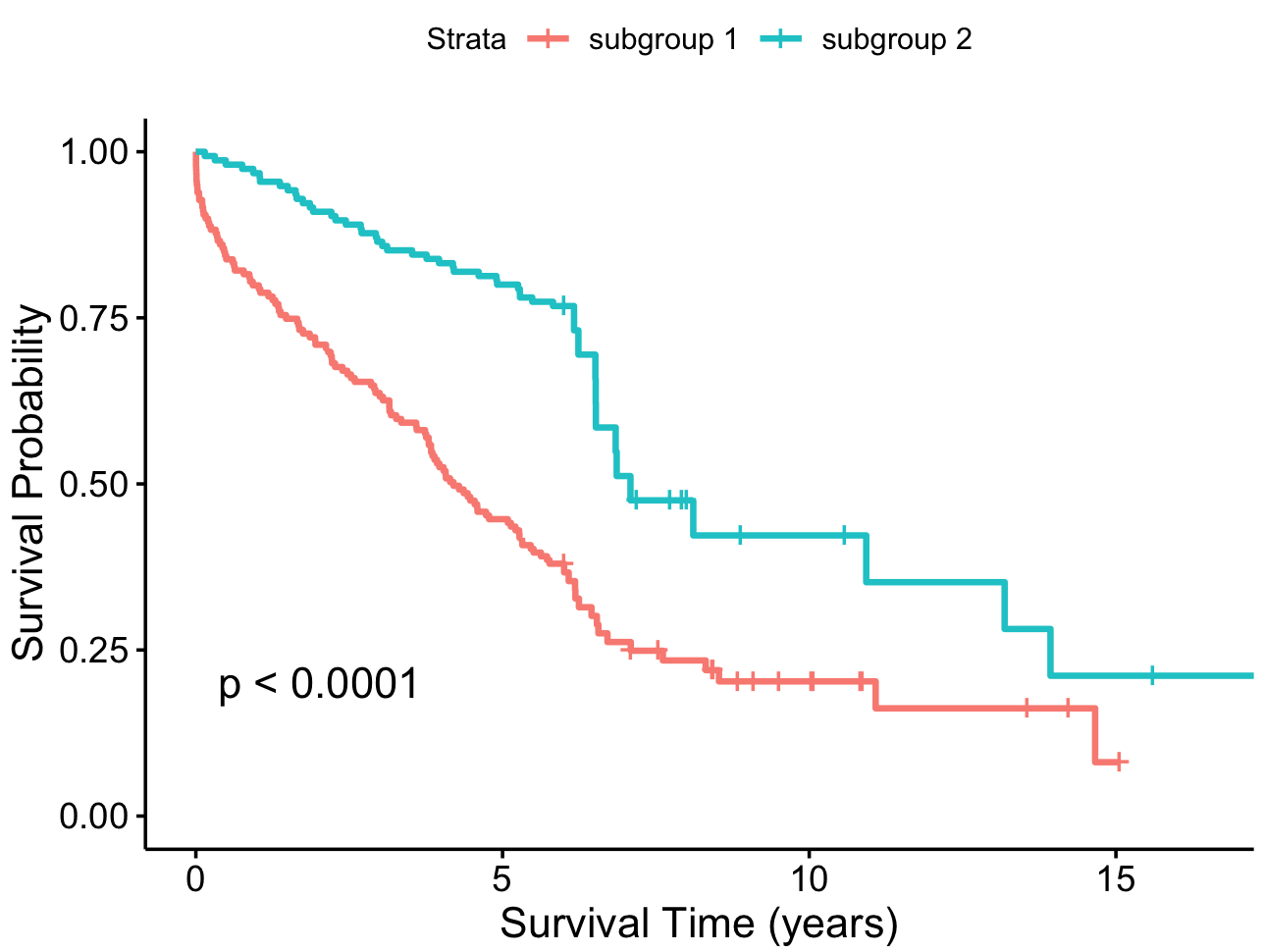}}\\
 LDA-(a) & LDA-(b) & LDA-(c) \\
 \raisebox{-.5\height}{\includegraphics[width=0.33\textwidth]{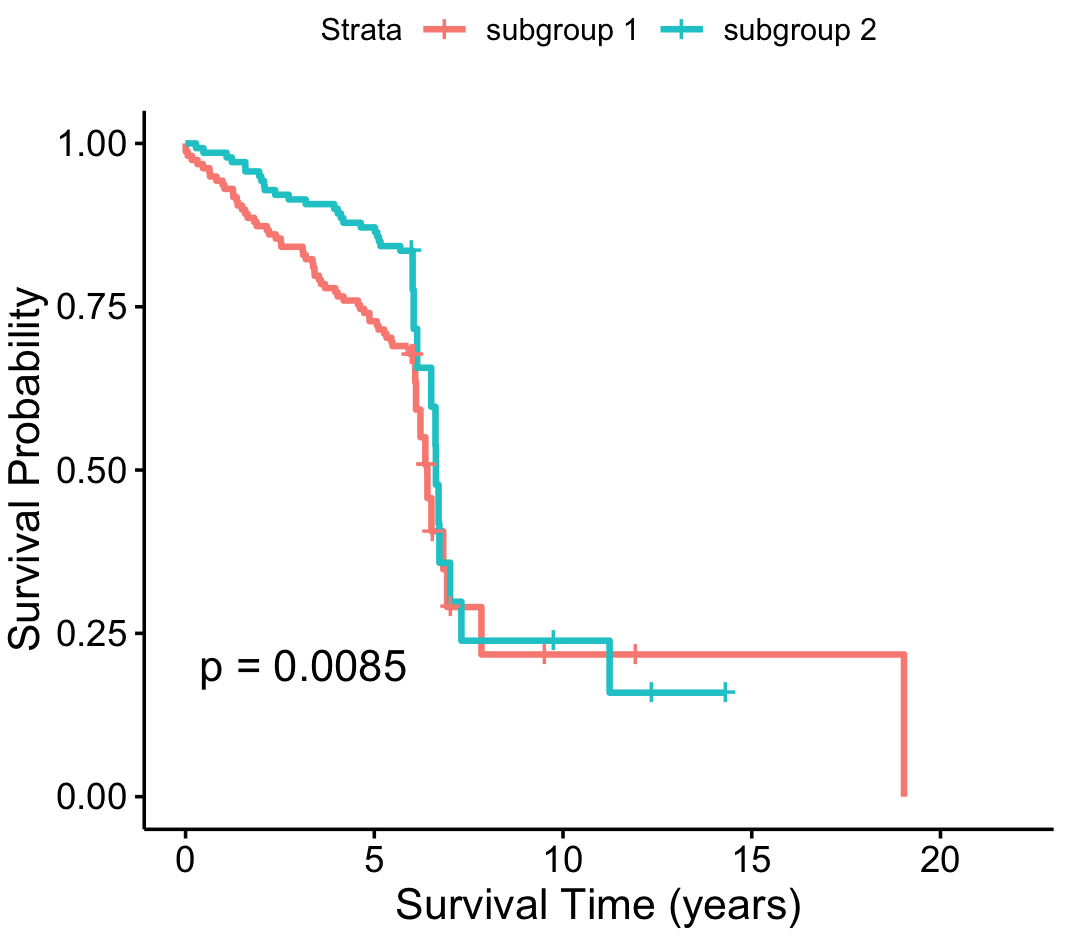}} & \raisebox{-.5\height}{\includegraphics[width=0.33\textwidth]{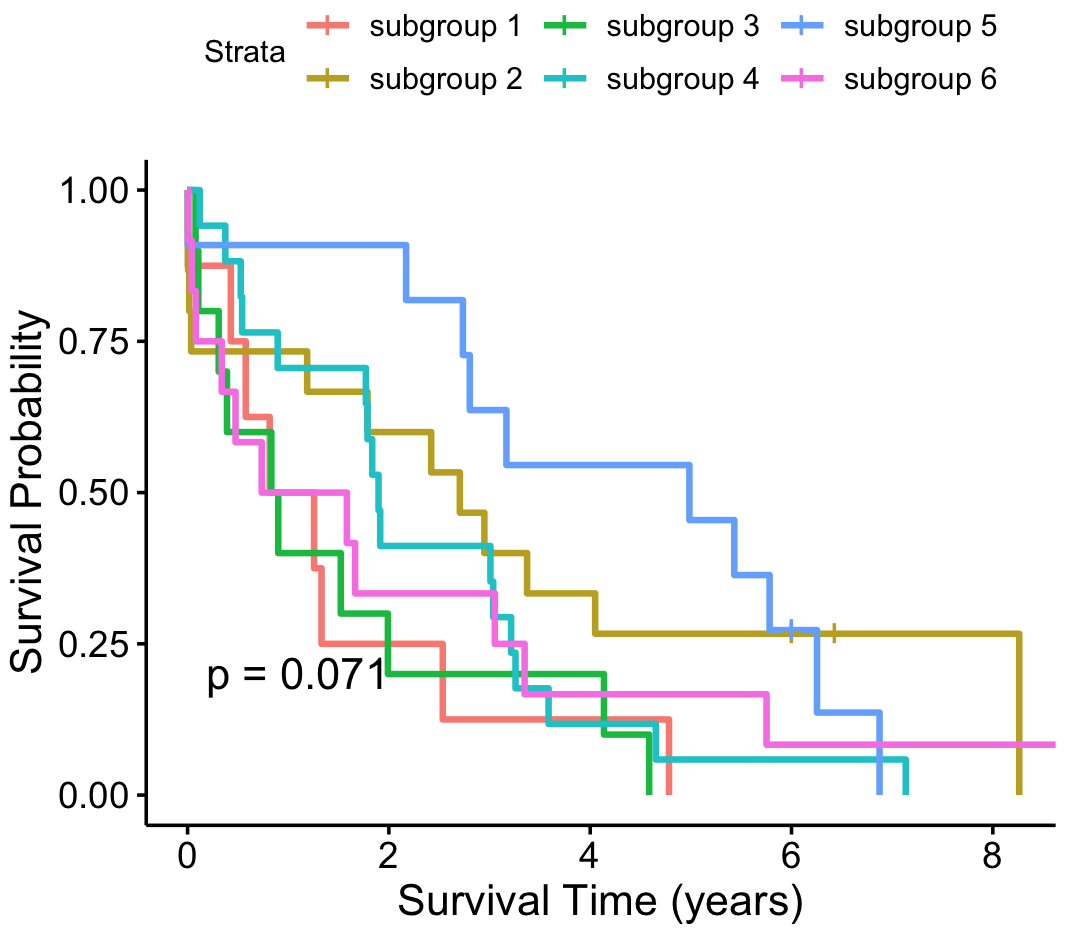}} & \raisebox{-.5\height}{\includegraphics[width=0.33\textwidth]{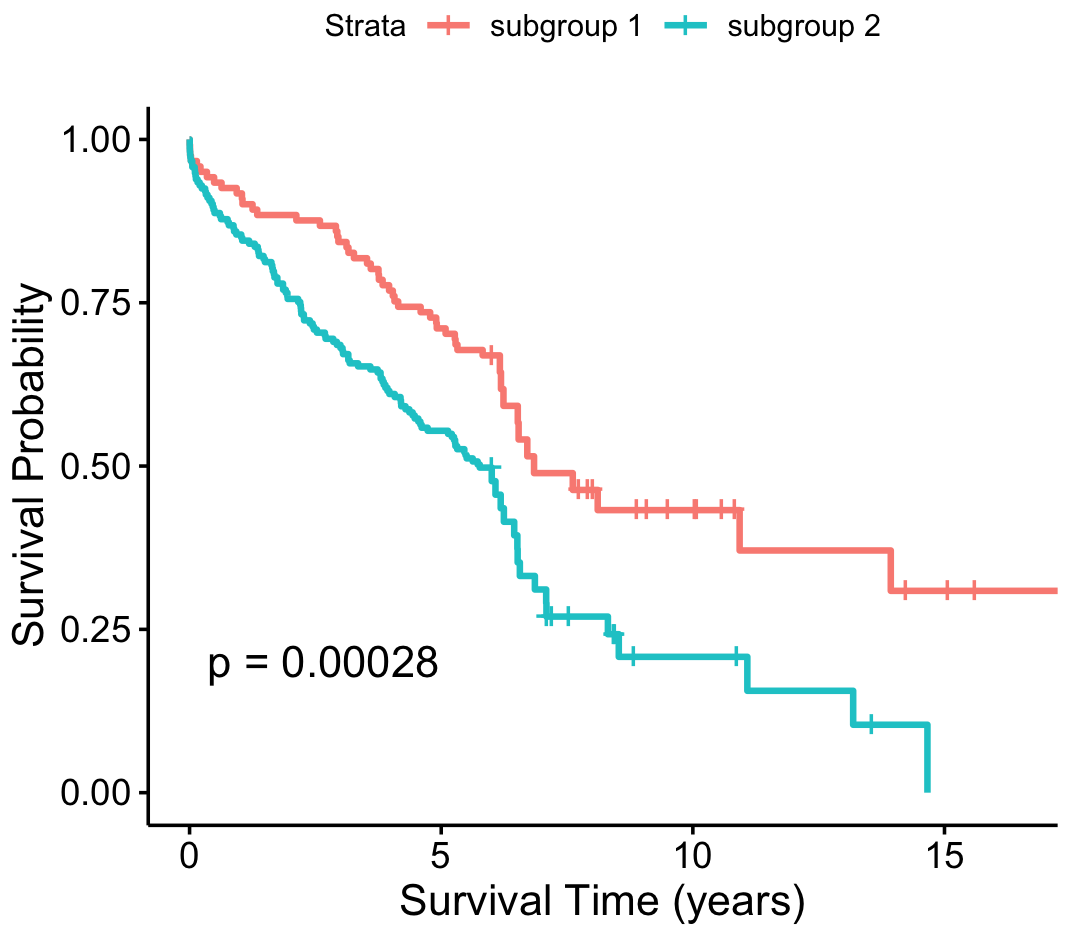}}\\
PDM-(a) & PDM-(b) & PDM-(c) \\
  \end{tabular}
  \caption{Survival analysis of the patient subgroups discovered by LDA and PDM from the Osteoporosis, Delirium/Dementia, and COPD/Bronchiectasis Cohorts. }
  \label{fig.surv}
\end{figure}

Kaplan-Meier survival curves of the selected patient subgroups discovered by LDA and PDM are depicted in Figure \ref{fig.surv} for three cohorts.

Survival analysis of patient subgroups discovered by LDA for the Osteoporosis Cohort depicted in Figure \ref{fig.surv} LDA-(a) showed significant difference between the survival curves of patient subgroups with p$<0.0001$. The patients in Subgroup 1 have a obviously distinguished worse survival rate than those in Subgroup 2. Thus, at the point of clinical care, the patients in Subgroup 1 should receive more attention than those in Subgroup2. Figure \ref{fig.surv} LDA-(b) shows significant difference at level $p<0.01$ between the survival curves of patient subgroups for the Delirium/Dementia Cohort. Three identified patient subgroups are distinguishable with disparate survival curves. The patients in Subgroups 2 have a better survival rate than those in Subgroups 1 and 3. The patients in Subgroups 1 have the lowest survival rate across the survival time distribution. Figure \ref{fig.surv} LDA-(c) shows a significant difference between the survival curves of patient subgroups with p$<0.0001$ for the COPD/Bronchiectasis Cohort. The patients in Subgroup 2 have a prominent lower risk and better survival than those in Subgroup 1. 

Figures \ref{fig.surv} PDM-(a), PDM-(b), and PDM-(c) show the survival analysis on the patient subgroups discovered by PDM for three cohorts. Figure \ref{fig.surv} PDM-(a) showed  difference between the survival curves of patient subgroups at the level of 0.01 for the Osteoporosis Cohort. The survival curves are similar to those by LDA. The patients in Subgroup 2 have a better survival rate than those in Subgroup 1 when survival time is approximately $<$ 6 years. The survival probability of both subgroups drops dramatically when survival time is between 6 years and 8 years. The survival probability of patients in Subgroup 2 decreases more rapid than those in Subgroup 1 after survival time $>$ 11 years. Though Figure \ref{fig.surv} PDM-(b) does not show significant difference between the survival curves of patient subgroups for the Dementia Cohort, patient subgroups are distinguishable with disparate survival curves. For example, the patients in Subgroups 1, 3, and 6 have similar survival curves when survival time is approximately $<$ 5 years. However, the patients in Subgroups 6 have longer survival time than those in Subgroups 1 and 3. The patients in Subgroups 2 and 5 have better survival than other subgroups across the survival time distribution, and those in Subgroup 2 have better survival than those in Subgroup 5 when survival time is $>$ 7 years. Figure \ref{fig.surv} PDM-(c) shows that the survival curves of patient subgroups are different at the level of <0.001 for the COPD Cohort. The patients in Subgroup 1 have a noticeable better survival rate than those in Subgroup 2. These results explicitly show the effectiveness of unsupervised machine learning models in stratifying patients into groups with different risks. 

\subsubsection{Statistical analysis}

\begin{sidewaystable}
    \centering
    \small
    \begin{tabular}{lccc|cccc}
    \hline
     Methods & \multicolumn{3}{c|}{LDA} & \multicolumn{4}{c}{PDM} \\
     \hline
    Subgroups &  Subgroup 1 & Subgroup 2 & p-value &  &  Subgroup 1 & Subgroup 2 & p-value \\
    \hline
    \# Patients (\%) & 271 (69.8\%) & 117 (30.2\%) &   &  & 216 (55.7\%) & 172 (44.3\%) &  \\
    \hline
    \textit{\textbf{Sex}} &  &  & 0.034 &  &   &  & 0.096 \\
    \# Male (\%) & 252 (93.0\%) & 115 (98.3\%) &   &  & 8 (3.7\%) & 13 (7.6\%) &   \\
    \# Female (\%) & 19 (7.0\%) & 2 (1.7\%) &   &  & 208 (96.3\%) & 159 (92.4\%) &   \\
    \hline
    Median Age & 78.3 & 66.3 & \textbf{$<$0.001}   &  & 71.1 & 74.6 & 0.002 \\
    \hline
    Median \# Diagnosis & 414.0 & 390.0 & 0.007  &  & 405 & 408 & 0.815  \\
    \hline
    Median ECI & 6.0 & 4.0 & \textbf{$<$0.001}   &  & 6.0 & 5.0 & 0.013  \\
    \hline
    \textit{\textbf{ECI Groups}} &  &  & \textbf{$<$0.001}  &   &  &  & 0.331 \\
    \# Patients in ECI (0-1) & 7 (2.6\%) & 9 (7.7\%)  &  &  & 9 (4.2\%) & 7 (4.1\%)   \\
    \# Patients in ECI (2-4) & 65 (24.0\%) & 64 (54.7\%)   &  &  & 65 (30.1\%) & 64 (37.2\%) \\
    \# Patients in ECI (5+) & 199 (73.4\%) & 44 (37.6\%) &   &  & 142 (65.7\%) & 101 (58.7\%) \\
    \hline
     \textit{\textbf{ECI Categories with} p$<0.001$} &  &  &  &  &  & \\
    Congestive heart failure & 48 (17.7\%) & 5 (4.3\%) & \textbf{$<$0.001} & Psychoses & 65 (30.1\%) & 23 (13.4\%) & \textbf{$<$0.001} \\ 
    Pulmonary circulation disorders & 28 (10.3\%) & 1 (0.9\%) & \textbf{$<$0.001} \\ 
    Hypertension & 209 (77.1\%) & 64 (54.7\%) & \textbf{$<$0.001} \\
    Other neurological disorders & 85 (31.4\%) & 16 (13.7\%) & \textbf{$<$0.001} \\
    Weight loss & 97 (35.8\%) & 16 (13.7\%) & \textbf{$<$0.001} \\
    Fluid and electrolyte disorders & 139 (51.3\%) & 30 (25.6\%) & \textbf{$<$0.001}\\
    Psychoses & 85 (31.4\%) & 3 (2.6\%) & \textbf{$<$0.001} \\
    \hline
    \end{tabular}
    \caption{Demographics and statistics of patient subgroups identified by LDA and PDM from the Osteoporosis Cohort. Statistically significance is based on the Kruskal-Wallis Test (p $<$ 0.001). The complete analysis for all ECI categories is provided in the supplemental material. }
    \label{tab.sub.ost}
\end{sidewaystable}

\begin{sidewaystable}
    \centering
    \small
    \begin{tabular}{l|cccc|cccccccc}
    \hline
    Methods & \multicolumn{4}{c|}{LDA} & \multicolumn{8}{c}{PDM} \\
     \hline
    Subgroups &  Subgroup 1 & Subgroup 2 & Subgroup 3  & p-value &  & Subgroup 1 & Subgroup 2 & Subgroup 3 & Subgroup 4 & Subgroup 5 & Subgroup 6 & p-value  \\
    \hline
    \# Patients (\%) & 167 (55.0\%) & 72 (23.7\%) & 65 (21.3\%) &   &   & 47 (15.5\%) & 41 (13.5\%) & 34 (11.2\%) & 62 (20.4\%) & 56 (18.4\%) & 64 (21.1\%) \\
    \hline
    \textit{\textbf{Sex}} &  &  &  & 0.114 &  &  &  &  &  &  & 0.445 \\
    \# Male (\%) & 44 (26.3\%) & 28 (38.9\%) & 23 (35.4\%)  &  &  & 16 (34.0\%) & 15 (36.6\%) & 9 (26.5\%) & 19 (30.6\%) & 12 (21.4\%) & 24 (37.5\%)  &  \\
    \# Female (\%) & 123 (73.7\%) & 44 (61.1\%) & 42 (64.6\%) &  &  & 31 (66.0\%) & 26 (63.4\%) & 25 (73.5\%) & 43 (69.4\%) & 44 (78.6\%) & 40 (62.5\%) &  \\
    \hline
    Median Age & 86.4 & 79.7 & 82.9 & \textbf{$<$0.001} &  &  82.4 & 84.0 & 84.9 & 84.3 & 85.7 & 82.3 & 0.623\\
    \hline
    Median \# Diagnosis & 385.0 & 414.0 & 376.0 & 0.009 & &   383.0 & 399.0 & 385.5 & 409.0 & 411.5 & 368.5 & 0.007\\
    \hline
    Median ECI & 8.0 & 8.0 & 7.0 & \textbf{$<$0.001} &  &  8.0 & 8.0 & 7.0 & 8.0 & 8.0 & 8.0 & 0.130  \\
    \hline
    \textit{\textbf{ECI Groups} }&  &  &  & \textbf{$<$0.001} &  &  &  &  &  &  & 0.874 \\
    \# Patients in ECI (0-1) & 0 (0.0\%) & 0 (0.0\%) & 0 (0.0\%) &  &  & 0 (0.0\%) & 0 (0.0\%) & 0 (0.0\%) & 0 (0.0\%) & 0 (0.0\%) & 0 (0.0\%) & \\
    \# Patients in ECI (2-4) & 5 (3.0\%) & 2 (2.8\%) & 12 (18.5\%) &  &  & 2 (4.3\%) & 2 (4.9\%) & 2 (5.9\%) & 3 (4.8\%) & 4 (7.1\%) & 6 (9.4\%) & \\
    \# Patients in ECI (5+) & 162 (97.0\%) & 70 (97.2\%) & 53 (81.5\%) &   &  & 45 (95.7\%) & 39 (95.1\%) & 32 (94.1\%) & 59 (95.2\%) & 52 (92.9\%) & 58 (90.6\%)&\\
    \hline
     \textit{\textbf{ECI Categories with} p$<0.001$} &  &  &   &  \\
    Hypertension & 140 (83.8\%) & 58 (80.6\%) & 39 (60.0\%) & \textbf{$<$0.001} \\ 
    \hline
    \end{tabular}
    \caption{Demographics and statistics of patient subgroups identified by LDA and PDM from the Delirium/Dementia Cohort. Statistically significance is based on the Kruskal-Wallis Test (p $<$ 0.001). The complete analysis for all ECI categories is provided in the supplemental file.}
    \label{tab.sub.dem}
\end{sidewaystable}

\begin{sidewaystable}
    \centering
    \small
    \begin{tabular}{l|ccc|p{1.5cm}ccc}
    \hline
    Methods & \multicolumn{3}{c|}{LDA} & \multicolumn{4}{c}{PDM} \\
    \hline
    &  Subgroup 1 & Subgroup 2 & p-value &  &  Subgroup 1 & Subgroup 2 & p-value \\
    \hline
    \# Patients (\%) & 495 (72.3\%) & 190 (27.7\%) &  & & 207 (30.2\%) & 478 (69.8\%) &  \\
    \hline
    \textit{\textbf{Sex}} &  &  & 0.005 &   & &  & 0.489 \\
    \# Male (\%) & 260 (52.5\%) & 77 (40.5\%) &   & & 106 (51.2\%) & 231 (48.3\%) &   \\
    \# Female (\%) & 235 (47.5\%) & 113 (59.5\%)) &   & & 101 (48.8\%) & 247 (51.7\%) & \\
    \hline
    Median Age & 76.0 & 66.8 & \textbf{$<$0.001}  & & 71.1 & 74.6 & 0.002 \\
    \hline
    Median \# Diagnosis & 403.0 & 402.0 & 0.791   & & 409.0 & 399.5 & 0.526\\
    \hline
    Median ECI & 9.0 & 6.0 & \textbf{$<$0.001}   & & 7.0 & 8.0 & \textbf{$<$0.001} \\
    \hline
    \textit{\textbf{ECI Groups}} &  &  & \textbf{$<$0.001} &  &  &  & 0.016   \\
    \# Patients in ECI (0-1) & 0 (0.0\%) & 2 (1.1\%)   &  & & 2 (1.0\%) & 0 (0.0\%)  \\
    \# Patients in ECI (2-4) & 15 (3.0\%) & 35 (18.4\%))  &  & & 21 (10.1\%) & 29 (6.1\%) \\
    \# Patients in ECI (5+) & 480 (97.0\%) & 153 (80.5\%)  &  & & 184 (88.9\%) & 449 (93.9\%)\\
    \hline
     \textit{\textbf{ECI Categories with} p$<0.001$} &  &  &  &  &  &  \\
    Congestive heart failure & 264 (53.3\%) & 27 (14.2\%) & \textbf{$<$0.001} & Fluid and electrolyte disorders & 111 (53.6\%) & 337 (70.5\%)) & \textbf{$<$0.001} \\ 
    Cardiac arrhythmias & 350 (70.7\%) & 102 (53.7\%) & \textbf{$<$0.001} \\ 
    Pulmonary circulation disorders & 154 (31.1\%) & 23 (12.1\%) & \textbf{$<$0.001} \\
    Renal failure & 123 (24.8\%) & 19 (10.0\%) & \textbf{$<$0.001} \\
    Obesity & 75 (15.2\%) & 50 (26.3\%) & \textbf{$<$0.001} \\
    Weight loss & 191 (38.6\%) & 37 (19.5\%) & \textbf{$<$0.001} \\
    Fluid and electrolyte disorders & 380 (76.8\%) & 68 (35.8\%) & \textbf{$<$0.001}\\
    Deficiency anaemia & 110 (22.2\%) & 21 (11.1\%) & \textbf{$<$0.001} \\
    Psychoses & 140 (28.3\%) & 19 (10.0\%) & \textbf{$<$0.001} \\
    \hline
    \end{tabular}
    \caption{Demographics and statistics of patient subgroups identified by LDA and PDM from the COPD/Bronchiectasis Cohort. Statistically significance is based on the Kruskal-Wallis Test (p $<$ 0.001). The complete analysis for all ECI categories is provided in the supplemental file.}
    \label{tab.sub.copd}
\end{sidewaystable}

Tables \ref{tab.sub.ost}, \ref{tab.sub.dem}, and \ref{tab.sub.copd} list the statistics of demographics, number of diagnoses, and ECI scores for the patient subgroups identified by LDA and PDM for the Osteoporosis, Delirium/Dementia, and COPD/Bronchiectasis Cohorts, respectively. Statistically significance is based on the Kruskal-Wallis Test (p $<$ 0.001). Reported also includes the ECI categories that are different among patient subgroups with statistical significance (p-value$<0.001$) in each cohort. The complete analysis for each ECI category can be found in the supplemental material.

We first analyze the patient subgroups identified by LDA. As shown in Table \ref{tab.sub.ost} for the Osteoporosis Cohort, the difference of sex in two patient subgroups is not significantly different. The patients in Subgroup 1 are older than those in Subgroup 2 with statistically significance at the level of 0.001, which might be the reason that the patients of Subgroup 1 have worse survival rate in Figure \ref{fig.surv} (a).  Median ECI scores between two subgroups are statistically different at the level of 0.001. The patients in Subgroup 1 have a higher median ECI score, which means they have more comorbidities. 73.4\% of patients in Subgroup 1 also have a larger number of comorbidites in terms of ECI (5+). The result is consistent with their survival analysis in Figure \ref{fig.surv} (a). Seven ECI categories, including congestive heart failure, pulmonary circulation disorders, hypertension, other neurological disorders, weight loss, fluid and electrolyte disorders, and psychoses, are statistically different between two patient subgroups at the level of 0.001. The diseases in these ECI categories potentially contributed to differentiate patients into subgroups for the Osteoporosis Cohort. 

For the patient subgroups identified by LDA from the Delirium/Dementia Cohort, there is statistical significant difference in age but not in sex among three subgroups. The fact that Subgroup 2 has better survival than Subgroup 1 is mainly due to the younger age of Subgroup 2. This result is consistent with previous findings that age is a strong risk factor for dementia \cite{gao1998relationships}. No statistical significance is found in the number of diagnoses at the level of 0.001. The patients in Subgroup 1 and those in Subgroup 2 have a similar number of comorbidities in terms of median number of ECI and ECI (5+). Hypertension is the ECI category that has statistically significant difference between three subgroups, which is also consistent with the outcome of large observational studies that that hypertension plays a role in dementia and Alzheimer's disease \cite{tzourio2007hypertension}.

There is also statistical significant different in age but not in sex for the patient subgroups identified by LDA from the COPD/Bronchiectasis Cohort. The patients in Subgroup 1 have a higher median ECI score and a much larger number of comorbidities in terms of ECI (5+) than those in Subgroup 2 with a statistically significant difference at p$<0.001$. This result is consistent with the survival analysis in Figure \ref{fig.surv} (c). Nine ECI categories are statistically different between two patient subgroups at the level of 0.001. These categories include congestive heart failure, cardiac arrhythmias, pulmonary circulation disorders, renal failure, obesity, weight loss, fluid and electrolyte disorders, deficiency anaemia, and psychoses. Interestingly, obesity appears to be associated with better outcome (15.2\% in Subgroup 1 and 26.3\% in Subgroup 2) and weight loss with worse outcome (38.6\% in Subgroup 1 and 19.5\% in Subgroup 2). That is likely related to the fact that being underweight in COPD/bronchiectasis is often a bad prognostic factor since the respiratory muscles lose strength during severe weight loss, which leads to respiratory failure. Being extremely obese also worsens COPD/bronchiectasis, but is not addressed in our analysis. As expected, psychosis are the comorbidities that worse COPD/bronchiectasis associated with the outcomes.

The first observation from the results of patient subgroups identified by PDM is that, unlike LDA, PDM does not differentiate patients into subgroups based on age and sex. No statistical significance are found in age and sex for the patient subgroups of three cohorts. This result validates the ability of PDM in removing the factors of age and sex for discovering patient subgroups, which enables analysis of hidden patterns of diseases that are of greater interest in epidemiology research. For the Osteoporosis Cohort, no statistical significance is found in the number of diagnoses, as well as the number of comorbidities. Only one ECI category, i.e., psychoses, is statistically different between two patient subgroups at the level of 0.001. For the Delirium/Dementia Cohort, no ECI category was found different with statistical significance among patient subgroups. This analysis could not find reasons why Subgroup 5 has the best survival according to the survival curves in Figure \ref{fig.surv} PDM-(b). This result implies that the conventional statistical analysis could not identify significant factors that cause differential patient subgroups discovered by PDM. In other words, PDM might leverage other undiscovered hidden disease patterns to discover these patient subgroups. For the COPD/Bronchiectasis Cohort, the patients in Subgroup 2 are older than those in Subgroup 1 but without statistical significance at the level of 0.001. ECI scores between two subgroups are statistically different with p$<0.001$. This result, similar to that of LDA, indicates that age portends worse prognosis as does greater comorbidities (ECI). Only one ECI categories, i.e., fluid and electrolyte disorders, are statistically different between two patient subgroups at the level of 0.001.

\section{Discussion and Conclusion}

In this study, we investigate the applications of unsupervised machine learning approaches in discovering latent disease clusters and patient subgroups using the EHR data. We utilized LDA, an unsupervised probabilistic generative model in the rubric of topic models, and proposed a novel unsupervised machine learning approach, named PDM. PDM extends the conventional LDA and uses a Poisson distribution to model patients' disease diagnoses and to alleviate age and sex factors by considering both observed and expected observations. We applied LDA and PDM to two clinical use cases: discovery of latent disease clusters and patient subgroups using EHRs. 

Both approaches were evaluated on the diagnostic EHR data of three cohorts, namely the Osteoporosis Cohort, the Delirium/Dementia Cohort, and the COPD/Bronchiectasis Cohort, retrieved from the REP medical linkage system. We verified the effectiveness of discovering latent disease clusters through the visualization of disease representation in the latent topic space. The 2-D scattered plot showed that PDM discovered explicitly dichotomized disease clusters based on the latent patterns hidden in the EHR data than LDA. This result implies that we could utilize these disease clusters to identify multiple latent comorbities, which could be used to calculate excess risk above what would be expected for a given age and sex.   

Furthermore, we applied LDA and PDM to discover patient subgroups, and carried out survival analysis on these subgroups. The experimental results show that LDA could stratify patients into more differentiable subgroups than PDM in terms of p-values. However, those subgroups identified by LDA are highly associated with patients' age and sex. Though the difference between the subgroups discovered by PDM has worse p-values, these patient subgroups might imply the underlying patterns of diseases of greater interest in epidemiology research by alleviating the impact of age and sex. Therefore, the proposed PDM might be a better option than LDA for studying latent disease patterns in aging cohorts for which we would like to alleviate the impact of age and sex since they are major drivers of aging-associated diseases.

Due to the similarity of unsupervised machine learning and human learning, unsupervised learning is more closely aligned with artificial intelligence (AI), where a computer is expected to learn to identify complex processes and patterns without a human's guidance. Compared to the supervised machine learning models that lack of generalizability and suffer from infeasibility of discovering novel patterns from EHRs~\cite{miotto2016deep}, the unsupervised machine learning techniques utilized to discover particular comorbidity clusters from EHRs that may reflect underlying mechanisms (``latent traits'') would help define new domains of risk. The disease clusters discovered by PDM might contain new potential risks for  diseases. Moreover, unsupervised machine learning approaches could be used to stratify patients into subgroups with similar characteristics and risks. Discovering differentiated patient subgroups will not only facilitate epidemiological analysis and research, but enable personalized care that will improve the efficiency and effectiveness of disease prevention, diagnosis, and treatment. We believe that both approaches could be leveraged to create an AI platform for exploiting the rich data resources of the REP, and likewise be served as a model for use by others with EHRs.

The limitations of this study are four-fold. First, LDA and the proposed PDM model were tested on cohorts with a relative small number of patients and diagnoses due to the computational cost of MH algorithm. Faster MCMC methods should be considered in future work so that the proposed model could be scaled up to larger cohorts. Second, the time stamp of diagnosis, which is an important feature for epidemiology, was not considered in LDA and PDM models. Third, PDM was only evaluated on the EHR data from the REP. In the future, we will evaluate the generalizability of the proposed model on the EHR data from other institutions. Fourth, unsupervised machine learning approaches other than probabilistic generative models, such as deep neural network based models \cite{miotto2016deep,choi2016doctor,choi2017gram}, were not considered in this study. In the future, we will compare the proposed approaches with the deep neural network based models in discovering latent disease clusters and patient subgroups using EHRs.


\section*{Acknowledgment}
\addcontentsline{toc}{section}{Acknowledgment}
This work was supported by NIH grants P01AG004875, R01GM102282, U01TR002062, and R01LM011934, and made possible by the Rochester Epidemiology Project (R01AG034676) and the U.S. Public Health Service. 

\makeatletter
\renewcommand{\@biblabel}[1]{\hfill #1.}
\makeatother

\bibliographystyle{IEEEtran}
\bibliography{pdm}

\end{document}